\documentclass[%
aip,
jcp,%
amsmath,amssymb,
reprint%
]{revtex4-1}
\usepackage[latin1]{inputenc}

\usepackage{amsmath}
\usepackage{amsfonts}
\usepackage{amssymb,bm}
\usepackage{graphicx}
\usepackage{physics}
\usepackage{upgreek}
\usepackage[outercaption]{sidecap}   
\usepackage{color}
\usepackage[normalem]{ulem}
\usepackage{txfonts}

\DeclareMathAlphabet{\pazocal}{OMS}{zplm}{m}{n}

\begin{document}

\def\bra#1{\left<{#1}\right|}
\def\ket#1{\left|{#1}\right>}
\def\expval#1#2{\bra{#2} {#1} \ket{#2}}
\def\mapright#1{\smash{\mathop{\longrightarrow}\limits^{_{_{\phantom{X}}}{#1}_{_{\phantom{X}}}}}}

\title{A general non-adiabatic quantum instanton approximation}
\author{Joseph E. Lawrence}
\affiliation{Department of Chemistry, University of Oxford, Physical and Theoretical\\ Chemistry Laboratory, South Parks Road, Oxford, OX1 3QZ, UK}
\email{joseph.lawrence@chem.ox.ac.uk}
\author{David E. Manolopoulos}
\affiliation{Department of Chemistry, University of Oxford, Physical and Theoretical\\ Chemistry Laboratory, South Parks Road, Oxford, OX1 3QZ, UK}

\begin{abstract}
We present a general quantum instanton approach to calculating reaction rates for systems with two electronic states and arbitrary values of the electronic coupling. This new approach, which we call the non-adiabatic quantum instanton (NAQI) approximation, reduces to Wolynes theory in the golden rule limit and to a recently proposed projected quantum instanton (PQI) method in the adiabatic limit. As in both of these earlier theories, the NAQI approach is based on making a saddle point approximation to the time integral of a reactive flux autocorrelation function, although with a generalised definition of the projection operator onto the product states. We illustrate the accuracy of the approach by comparison with exact rates for one dimensional scattering problems and discuss its applicability to more complex reactions. 
\end{abstract}

\maketitle
\section{Introduction}

Rate processes which involve both nuclear quantum effects and the breakdown of the Born-Oppenheimer approximation are important in a wide variety of contexts, ranging from device physics to biology.\cite{Hammes-Schiffer10,Blumberger15,Oberhofer17,Dou18} For systems with more than a few degrees of freedom, exact wavefunction based approaches are impractical due to the exponential scaling of quantum mechanics with dimensionality. Imaginary time path integral techniques,\cite{Feynman65} which can accurately capture zero point energy and tunnelling effects and yet scale only linearly with system size, have therefore become popular for studying more complex reactions.  While there now exist several such methods which are routinely used to study electronically adiabatic reactions,\cite{Voth89,Miller03,Vanicek05,Craig05a,Craig05b,Collepardo08,Boekelheide11}  as well as theories which can be applied to electronically non-adiabatic reactions in the golden rule limit,\cite{Wolynes87,Zheng89,Bader90,Zheng91,Lawrence18,Thapa19,Fang19,Fang20} the development of accurate path integral techniques for more general electronically non-adiabatic reactions with intermediate electronic coupling strengths is still a very active area of research.\cite{Shushkov12,Richardson13,Ananth13,Duke16,Menzeleev14,Kretchmer16,Chowdhury17,Kretchmer18,Tao18,Tao19,Lawrence19b,Lawrence19a,Lawrence20}

When developing new path integral methods for studying non-adiabatic systems a common approach has been to generalise a pre-existing adiabatic method. One of the most successful path integral techniques for electronically adiabatic reactions is ring polymer molecular dynamics\cite{Craig04,Habershon13} (RPMD) reaction rate theory,\cite{Craig05a,Craig05b} and because of this much of the work in this area has focussed on trying to extend it to treat non-adiabatic systems.\cite{Shushkov12,Richardson13,Ananth13,Duke16,Menzeleev14,Kretchmer16,Chowdhury17,Kretchmer18,Tao18,Tao19,Lawrence19b} RPMD is particularly effective in the deep tunnelling regime because of its connection with the semiclassical instanton approximation,\cite{Richardson09} which uses a periodic imaginary time trajectory through the reaction barrier to describe the tunnelling process.\cite{Miller75} While it is not as generally applicable as RPMD, the semiclassical instanton formula is known to provide a highly accurate description of tunnelling in situations where there is a single dominant tunnelling path. Early work extending the semiclassical instanton approach to treat non-adiabatic reactions was based on assuming that the ``Im-F'' premise\cite{Chapman75,Callan77} could be applied to non-adiabatic systems, and succeeded in providing a theory which bridged between the golden rule and Born-Oppenheimer limits \cite{Cao95,Cao97,Schwieters98,Schwieters99} More recently, Richardson \emph{et al.}\cite{Richardson15a,Richardson15b,Heller20} have provided a rigorous derivation of the semiclassical instanton rate in the golden rule limit, and found some important differences between the resulting expression and that given by the Im-F formulation. However, their derivation has yet to be extended beyond the golden rule limit so that it can be applied to reactions with arbitrary electronic coupling strengths. 

In this paper we shall focus on another well known method, the quantum instanton approximation. Unlike the other methods discussed above, this approximation was in fact first suggested in the golden rule context by Wolynes\cite{Wolynes87} in 1987 (leading to what is now typically referred to as Wolynes theory), before the adiabatic counterpart was suggested by Miller \emph{et al.}\cite{Miller03} in 2003 (who gave it the name quantum instanton). Wolynes theory and the quantum instanton are both closely related to the semiclassical instanton, and all three can be interpreted as steepest descent approximations to the flux-flux correlation function expression for the reaction rate. However, whereas the semiclassical instanton simultaneously approximates integrals over both position and time, Wolynes theory and the quantum instanton involve just a single steepest descent approximation to the time integral. The resulting expressions only involve time-independent quantities, which can be evaluated by sampling imaginary time paths. 

Recently Vaillant \emph{et al.}\cite{Vaillant19} have suggested a slight modification of the original adiabatic quantum instanton which they have called the projected quantum instanton (PQI). This enforces sampling of paths close to the semiclassical instanton and results in an expression that is even more closely related to Wolynes theory. In the following we shall present a generalised approach to electronically non-adiabatic reactions that is applicable to arbitrary electronic coupling strengths between the golden rule and adiabatic limits and which reduces to Wolynes theory and the PQI approximation in these two limits, respectively.

We begin in Section~\ref{Exact_Theories} by discussing how the choice of the projection operators that are used to define the reactants and products affects the functional form of the reactive flux-flux correlation function of an electronically non-adiabatic reaction. Motivated by this discussion, we introduce a simple projection operator onto the product states which can be tuned so as to minimise the recrossing of the transition state dividing surface for any given non-adiabatic reaction. In Section~\ref{Theory}, we summarise the Wolynes theory and quantum instanton approaches to the golden rule and Born-Oppenheimer limits, before introducing a more general approach for the calculation of reaction rates for arbitrary electronic coupling strengths, which we shall call the non-adiabatic quantum instanton (NAQI) approximation. In Section~\ref{Results} we investigate the accuracy of the NAQI formula for a series of simple one-dimensional scattering problems for which the exact quantum mechanical reaction rates can be computed for comparison. Section~\ref{conclusion} concludes the paper, discussing several possible applications of the NAQI approach and the scope for further theoretical developments.

\section{Exact Reaction Rate Theory} \label{Exact_Theories}

The Hamiltonian for a general two level system can be written in the diabatic representation as
\begin{equation}
\hat{H} = \hat{H}_0\dyad{0}{0}+\hat{H}_1\dyad{1}{1} + \Delta(\dyad{0}{1}+\dyad{1}{0}),
\end{equation}
where $\ket{0}$ and $\ket{1}$ are the two diabatic electronic states. Since it suffices for our present purposes, we shall restrict our attention to simple one-dimensional scattering problems of the form
\begin{equation}
\hat{H}_i = \frac{\hat{p}^2}{2m}+{V}_i(\hat{q}),
\end{equation} 
in which $V_i(q)$ is the diabatic potential on electronic state $\ket{i}$. We shall also assume that the electronic coupling $\Delta$ is a constant, independent of the nuclear configuration $q$ (the Condon approximation). Within this simple framework, the adiabatic potentials, $U_{\pm}(q)$, are
\begin{equation}
U_{\pm}(q)=\frac{V_0(q)+V_1(q)}{2}\pm\frac{1}{2}\sqrt{(V_0(q)-V_1(q))^2+4\Delta^2}.
\end{equation}
Everything we shall have to say can readily be generalised to treat more complex multi-dimensional reactions, and to include non-Condon effects. However, Eqs.~(1) to (3) are all we shall need to make the points we would like to make here.

The exact quantum mechanical thermal rate constant for the transition from reactants to products can be written in the form\cite{Yamamoto60,Miller83}  
\begin{equation}
k Q_r = \frac{1}{2} \int_{-\infty}^{\infty}  c_{\mathrm{ff}}(t)\, \mathrm{d}t, \label{exact_rate}
\end{equation}
where $Q_r=\sqrt{m/2\pi\beta\hbar^2}$ is the reactant partition function per unit length and
\begin{equation}
 c_{\mathrm{ff}}(t)=\tr[e^{-\beta\hat{H}/2}\hat{F}e^{-\beta\hat{H}/2}e^{+i\hat{H}t/\hbar}\hat{F}e^{-i\hat{H}t/\hbar}]
\end{equation}
is a reactive flux autocorrelation function (with $\beta=1/k_{\rm B}T$). The flux operator is the Heisenberg time derivative of the projection onto the products,
\begin{equation}
\hat{F} = \frac{i}{\hbar}[\hat{H},\hat{P}_p].
\end{equation}
This formulation applies equally well in both the adiabatic and non-adiabatic limits with appropriate definitions of the projection operator $\hat{P}_p$.

In the adiabatic limit where the Born-Oppenheimer approximation is valid, it is usual to define $\hat{P}_p$ as
\begin{equation}
\hat{P}_p = \theta(s(\hat{{q}})),
\end{equation}
where $\theta(x)$ is a Heaviside step function and $s(q)=0$ is a position space dividing surface between the reactants [$s(q)<0$] and products [$s(q)>0$]. The flux operator then becomes
\begin{equation}
\hat{F} = \frac{\hat{p}}{2m} \frac{\partial s}{\partial q} \delta(s(\hat{q})) +   \frac{\partial s}{\partial q}\delta(s(\hat{q})) \frac{\hat{p}}{2m}.
\end{equation}
In the golden rule limit where $\Delta\to0$, it is more usual to define the rate in terms of a transition between the diabatic states $\ket{0}$ and $\ket{1}$, which gives
\begin{equation}
\hat{P}_p = \dyad{1}{1},
\end{equation}
and
\begin{equation}
\hat{F}=\frac{i}{\hbar}\Delta(\dyad{0}{1}-\dyad{1}{0}).
\end{equation}

\begin{figure}[t]
 \resizebox{1.0\columnwidth}{!} {\includegraphics{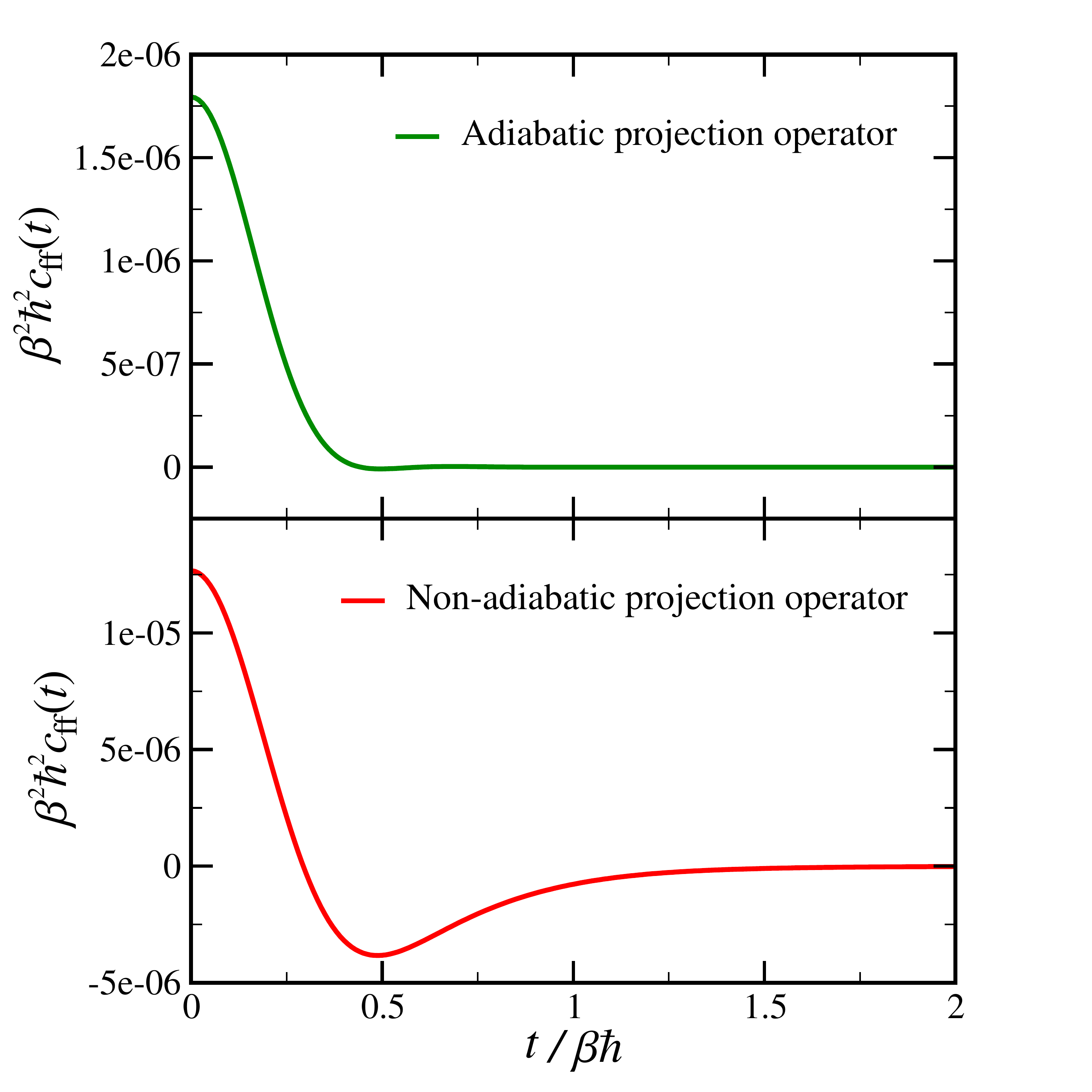}}
 \centering
 \caption{Flux-flux correlation functions for an exponential crossing model  in the adiabatic regime  (with $\beta A=48$, $mL^2/\beta\hbar^2=1/4$ and $\log_{10}(\beta\Delta)=1.5$).  Note the different scales on the y-axes in the two panels. The areas under both curves are the same. However the non-adiabatic projection operator clearly leads to a much longer-lived correlation function with a negative tail, which indicates recrossing of the dividing surface between reactants and products.\cite{Tromp87}}
 \label{SB_Rates}
 \end{figure}
\begin{figure}[t]
 \resizebox{1.0\columnwidth}{!} {\includegraphics{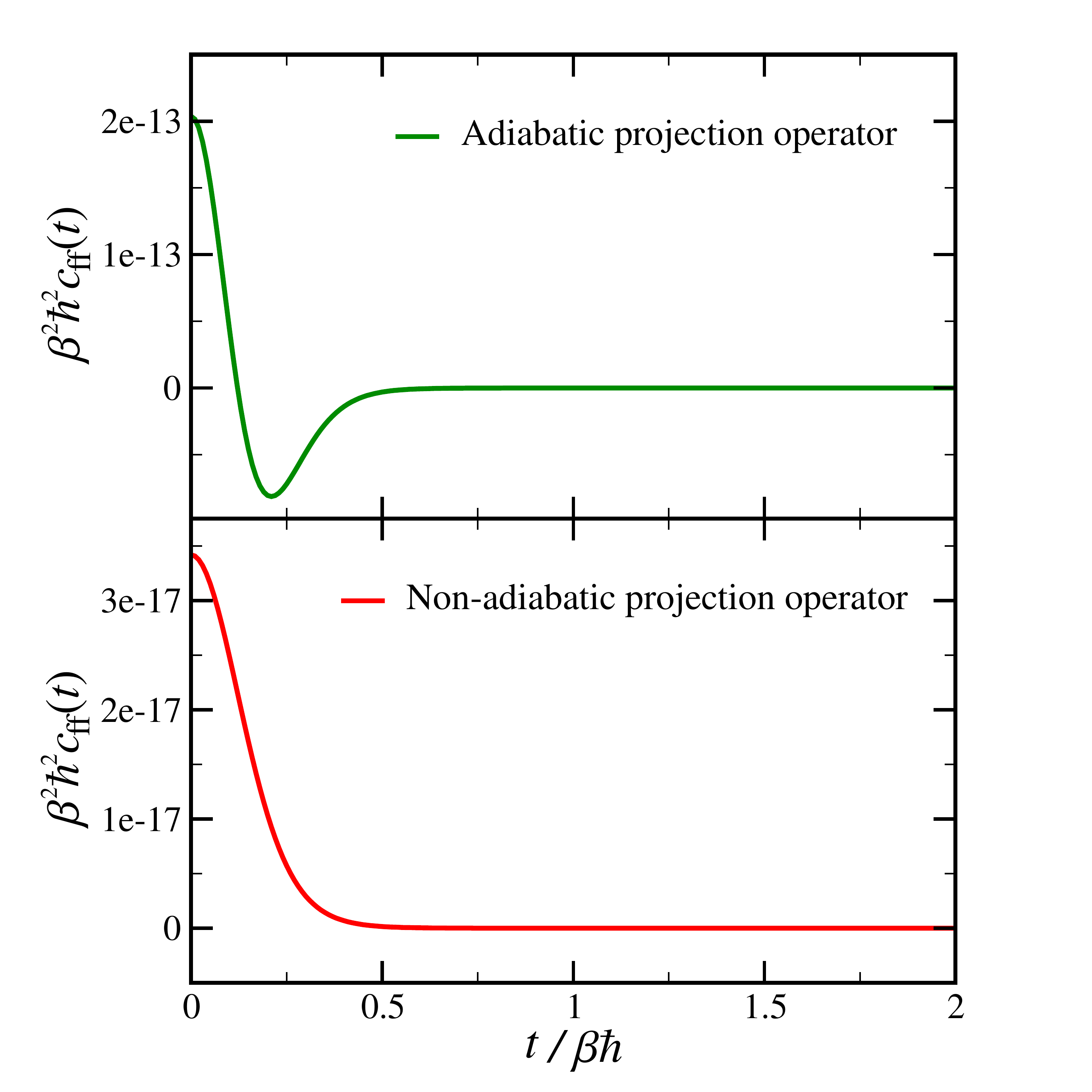}}
 \centering
 \caption{Flux-flux correlation functions for an exponential crossing model in the golden rule regime  (with $\beta A=48$, $mL^2/\beta\hbar^2=1/4$ and $\log_{10}(\beta\Delta)=-2$). Note the different scales on the y-axes in the two panels. The areas under both curves are the same. However the adiabatic projection operator clearly leads to a correlation function with a negative tail, which indicates recrossing of the dividing surface between reactants and products.\cite{Tromp87}}
 \label{SB_Rates}
 \end{figure}
 \begin{figure}[t]
 \resizebox{1.0\columnwidth}{!} {\includegraphics{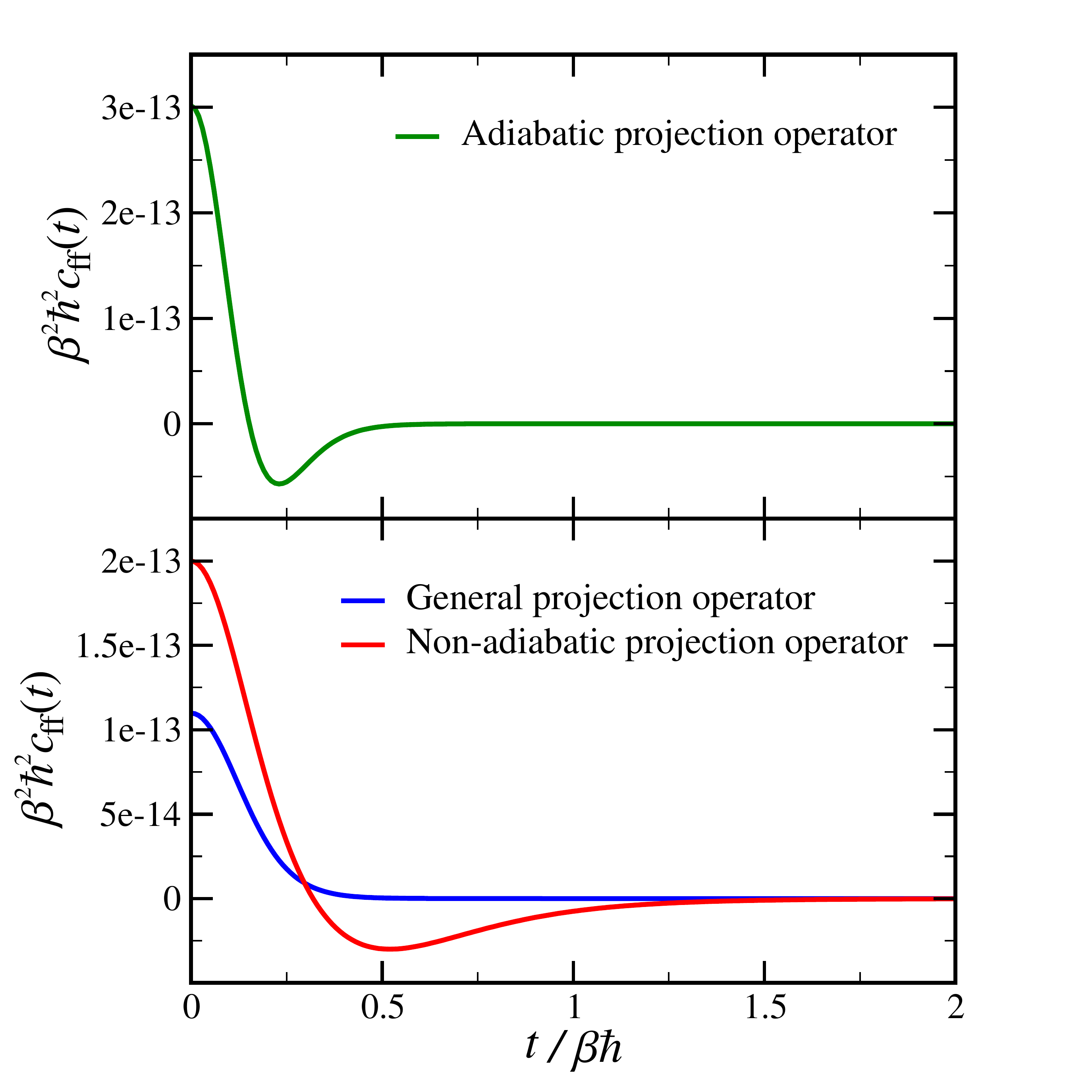}}
 \centering
 \caption{Flux-flux correlation functions for an exponential crossing model with parameters intermediate between the golden rule and the adiabatic limits ($\beta A=48$, $mL^2/\beta\hbar^2=1/4$ and $\log_{10}(\beta\Delta)=-0.25$). Note the different scales on the y-axes in the two panels. The plot clearly illustrates that in this regime the adiabatic projection operator leads to a correlation function with significant recrossing of the dividing surface and the non-adiabatic projection operator leads to a much longer lived correlation function with a negative tail. The generalised projection operator is seen to give a correlation function with almost no recrossing.}
 \label{SB_Rates}
 \end{figure}

In situations where the reactants and products can equally well be distinguished using either definition of $\hat{P}_p$, the rate constant is independent of the definition used. However, while the choice of projection operator does not change the rate, it does change the functional form of the flux-flux correlation function. To illustrate this we shall consider a simple model curve crossing problem 
\begin{subequations}
\begin{equation}
V_0(q) = A e^{+q/L}, 
\end{equation} 
\begin{equation}
V_1(q)=A e^{-q/L},
\end{equation} 
\end{subequations}
with $\beta A=48$, $mL^2/\beta\hbar^2=1/4$, and three different values of the electronic coupling strength ($\beta \Delta$).

Figure 1 shows the flux-flux correlation functions for this model problem with $\log_{10}(\beta\Delta)=1.5$ (approximately in the adiabatic limit), as calculated using both the adiabatic and non-adiabatic projection operators. The position space projection operator leads to a correlation function which decays quickly with almost no negative correlation, whereas the diabatic projection operator gives a correlation function that is much larger at $t=0$ and has a slowly decaying negative tail. The position space projection operator is thus seen to lead to far less recrossing than the diabatic projection operator in this regime. 

Figure 2 shows the flux-flux correlation function for $\log_{10}(\beta\Delta)=-2$, which is in the opposite (non-adiabatic) limit. In this case the amount of recrossing is reversed, with the diabatic projection operator leading to an approximately Gaussian flux-flux correlation function with minimal recrossing, whereas the position space projection operator gives rise to significant recrossing. Here the recrossing arises from the high probability that when the system passes through the dividing surface it will remain on the same diabatic surface, and hence will a short time later be reflected by the potential wall and return through the dividing surface.  

The projection operators in Eqs.~(7) and~(9) are clearly quite different. However we note that it is possible to recast the projection operator on to the products in the golden rule limit in a form that more closely resembles Eq.~(7),
\begin{equation}
\hat{P}_p= \theta(-\hat{\sigma}_z),
\end{equation}
where $\hat{\sigma}_z$ is the Pauli spin operator $\hat{\sigma}_z=\dyad{0}{0}-\dyad{1}{1}$. This suggests an obvious generalisation in which the argument of the Heaviside step function is taken to be a linear combination of $s(\hat{q})$ and $\hat{\sigma}_z$, 
\begin{equation}
\hat{P}_p(\alpha)=\theta(\cos(\alpha)s(\hat{q})-\sin(\alpha)\hat{\sigma}_z), \label{generalised_proj_op}
\end{equation}
or equivalently, provided $0\leq\alpha<\pi/2$, using the properties of the step function,
\begin{equation}
\hat{P}_p(\alpha)=\theta(s(\hat{q})-\tan(\alpha)\hat{\sigma}_z).
\end{equation}
This $\hat{P}_p(\alpha)$ is diagonal in the electronic basis with diagonal matrix elements
\begin{subequations}
\begin{equation}
\bra{0}\hat{P}_p(\alpha)\ket{0} = \theta(s(\hat{q})-\tan(\alpha)),
\end{equation}
\begin{equation}
\bra{1}\hat{P}_p(\alpha)\ket{1} = \theta(s(\hat{q})+\tan(\alpha)),
\end{equation}
\end{subequations}
and so we see that we can also think of the generalised projection operator as effectively giving rise to two separate position space dividing surfaces, one for each diabatic state.

In the Born-Oppenheimer limit it is clear that $\alpha\to0$, and in the non-adiabatic limit that $\alpha\to\pi/2$. In Figure 3 we show the flux-flux correlation functions for the system with $\log_{10}(\beta\Delta)=-0.25$, which is intermediate between the adiabatic and non-adiabatic limits. The results for all three projection operators are shown, with $s(q)=q/L$ and $\tan(\alpha)=12/5$ in the case of the generalised projection operator in Eq.~(14). We see that with this choice of $\alpha$ the correlation function is approximately Gaussian with minimal recrossing. In contrast both the purely position space and purely diabatic state projection operators have significantly higher initial values, along with regions of negative correlation corresponding to recrossing.

\section{Quantum Transition State Theories}\label{Theory}

The advantage of writing the rate in terms of the integral of a flux-flux correlation function with minimal recrossing is that one can then use this to develop a ``quantum-transition state theory''. By this we mean an approximate expression for the rate that only depends on time-independent quantities, such as Wolynes theory in the non-adiabatic limit\cite{Wolynes87} and the quantum instanton approximation in the adiabatic limit.\cite{Miller03} Here we first give a brief summary of these two existing quantum transition state theories, before showing how they can be generalised to give a method that is applicable to electronically non-adiabatic reactions with arbitrary electronic coupling strengths. 

\subsection{Wolynes Theory}

 In order to derive the Wolynes theory expression for the rate one begins by taking the golden rule limit of the flux-flux correlation function in Eq.~(5), with the flux operator defined as in Eq.~(10). Since $\bra{i}e^{-i\hat{H}t/\hbar}\ket{i}=e^{-i\hat{H}_it/\hbar}+\mathcal{O}(\Delta^2)$ and $\bra{i}e^{-i\hat{H}t/\hbar}\ket{j}=\mathcal{O}(\Delta)$ for $i\neq j$, it follows that in the golden rule ($\Delta\to0$) limit 
\begin{equation}
c_{\mathrm{ff}}(t) = \frac{2\Delta^2}{\hbar^2} \Re\big( c_{\mathrm{GR}}(t)\big),
\end{equation} 
where
\begin{equation}
c_{\mathrm{GR}}(t)= \tr_{\mathrm{n}}\left[e^{-\beta\hat{H}_0/2-i\hat{H}_0t/\hbar}e^{-\beta\hat{H}_1/2+i\hat{H}_1t/\hbar}\right]
\end{equation}
and $\tr_{\mathrm{n}}[\dots]$ denotes a trace over nuclear coordinates. Noting that introducing an arbitrary bias to products, $V_1(q)\to V_1(q)-\epsilon$, leads to
\begin{equation}
c_{\mathrm{GR}}(t)\to c_{\mathrm{GR}}(t)\,e^{+\beta\epsilon/2-i\epsilon t/\hbar}
\end{equation}
it is clear that changing the bias introduces an oscillatory component into the correlation function. In order to remove this oscillation and restore the approximately Gaussian behaviour seen for the symmetric problem in the lower panel of Fig.~2, one makes use of the relation, $c_{\mathrm{GR}}(-t)=c_{\mathrm{GR}}^{*}(t)$ to rewrite the rate in Eq.~(4) as
\begin{equation}
kQ_r = \frac{\Delta^2}{\hbar^2} \int_{-\infty}^{\infty} c_{\mathrm{GR}}(t)\, \mathrm{d}t, 
\end{equation}
and then performs the integration over time by shifting the contour of integration to pass through a saddle point of $c_{\mathrm{GR}}(t)$ on the imaginary time axis. This leads to the Wolynes theory approximation to the rate\cite{Wolynes87}  
\begin{equation}
k_{\mathrm{WT}} Q_r = \frac{\Delta^2}{\hbar} \sqrt{\frac{2\pi}{-\beta F_{\rm WT}''(\lambda_{\rm sp})}} e^{-\beta F_{\rm WT}(\lambda_{\rm sp})}, 
\end{equation}
where we have defined
\begin{align}
e^{-\beta F_{\rm WT}(\lambda)} &= \tr_\mathrm{n}\left[e^{-(\beta-\lambda)\hat{H}_0}e^{-\lambda\hat{H}_1}\right]\nonumber\\ &= c_{\mathrm{GR}}(i(\lambda-\beta/2)\hbar)
\end{align}
and the saddle point condition is $F_{\rm WT}'(\lambda_{\rm sp})=0$.

\subsection{Adiabatic Quantum Instanton}

In the adiabatic limit, for reactions which can be considered to proceed on the lower adiabatic surface, the flux-flux correlation function becomes
\begin{equation}
c_{\mathrm{ff}}(t)=\tr_{\mathrm{n}}\left[e^{-\beta\hat{H}_{\mathrm{BO}}/2}\hat{F}e^{-\beta\hat{H}_{\mathrm{BO}}/2}e^{+i\hat{H}_{\mathrm{BO}}t/\hbar}\hat{F}e^{-i\hat{H}_{\mathrm{BO}}t/\hbar}\right],
\end{equation}
where
\begin{equation}
\hat{H}_{\mathrm{BO}}= \frac{\hat{p}^2}{2m}+\hat{U}_-(q)
\end{equation}
and the flux operator is
\begin{equation}
\hat{F}=\frac{i}{\hbar}[\hat{H}_{\mathrm{BO}},\hat{P}_p].
\end{equation}
The original paper by Miller \emph{et al.}\cite{Miller03}~proposed two closely related ``quantum instanton" methods. The conceptually simpler of the two makes a second order cumulant approximation to the flux-flux correlation function, 
\begin{equation}
c_{\mathrm{ff}}(t) \simeq c_{\mathrm{ff}}(0)\exp(\frac{\ddot{c}_{\mathrm{ff}}(0)}{2c_{\mathrm{ff}}(0)}t^2),
\end{equation}
and then integrates over time to give
\begin{equation}
k_{\mathrm{QI}}Q_r = \sqrt{\frac{\pi c_{\mathrm{ff}}(0)}{-2\ddot{c}_{\mathrm{ff}}(0)}} c_{\mathrm{ff}}(0). \label{QI}
\end{equation}
The problem with this approach is that the second order cumulant expansion of the flux-flux correlation function can be a poor approximation for asymmetric reactions. This can be understood as arising for essentially the same reason as we have discussed above for Wolynes theory. Vaillant \emph{et al.}\cite{Vaillant19} have recently examined the problem in detail with a semiclassical analysis in which they showed that neither Eq.~(\ref{QI}) nor the alternative formulation of the quantum instanton method in the original paper by Miller \emph{et al.}\cite{Miller03} reduces to the semiclassical instanton approximation in the limit as $\hbar\to0$. 

In order to fix this problem, Vaillant {\em et al.} have suggested a modified method, the projected quantum instanton (PQI), in which the flux-flux correlation function is approximated as\cite{Vaillant19}
\begin{equation}
c_{\rm ff}(t) \simeq 2{\rm Re}\left[c_{\rm PQI}(t)\right],
\end{equation}
where
\begin{equation}
c_{\mathrm{PQI}}(t)= \tr_{\mathrm{n}}\left[U_r(i\beta\hbar/2-t)\,\hat{F}U_p(i\beta\hbar/2+t)\,\hat{F}\right],
\end{equation}
with
\begin{equation}
U_s(t)=e^{+i\hat{H}_{\mathrm{BO}}t/(2\hbar)}\hat{P}_se^{+i\hat{H}_{\mathrm{BO}}t/(2\hbar)}
\end{equation}
for $s=r\text{ and }p$, with $\hat{P}_r=\hat{1}-\hat{P}_p$.  The difference between Eqs.~(22) and (27) is that the exact $c_{\rm ff}(t)$ contains two additional terms of the form
\begin{equation*}
{\rm tr}_n\left[U_s(i\beta\hbar/2-t)\,\hat{F}U_s(i\beta\hbar/2+t)\,\hat{F}\right]
\end{equation*}
with $s=r$ and $p$. However, Vaillant {\em et al.}~argue that since the time integrals of these terms vanish in the semiclassical ($\hbar\to 0$) limit, they can safely be neglected when calculating the reaction rate.\cite{Vaillant19}

Substituting Eq.~(27) into Eq.~(4) and noting that $c_{\rm PQI}(-t)=c_{\rm PQI}^*(t)$ gives
\begin{equation}
kQ_r\simeq \int_{-\infty}^{\infty} c_{\rm PQI}(t)\,{\rm d}t.
\end{equation}
This time integral can be evaluated by steepest descent as in Wolynes theory to give
\begin{equation}
k_{\rm PQI}Q_r = {1\over\beta^2\hbar}\sqrt{2\pi\over -\beta F_{\rm PQI}''(\lambda_{\rm sp})}e^{-\beta F_{\rm PQI}(\lambda_{\rm sp})},
\end{equation}
where 
\begin{align}
e^{-\beta F_{\rm PQI}(\lambda)} &= (\beta\hbar)^2{\rm tr}_n\left[U_r(i(\beta-\lambda)\hbar)\,\hat{F}U_{p}(i\lambda\hbar)\,\hat{F}\right]\nonumber\\ &= (\beta\hbar)^2c_{\rm PQI}(i(\lambda-\beta/2)\hbar),
\end{align}
in which the factor of $(\beta\hbar)^2$ has been introduced to ensure dimensional consistency in Eq.~(32) and then compensated for in Eq.~(31). The saddle point condition is now $F_{\rm PQI}'(\lambda_{\rm sp})=0$, which is satisfied by the value of $\lambda$ that maximises $F_{\rm PQI}(\lambda)$ and minimises $c_{\rm PQI}(i(\lambda-\beta/2)\hbar)$.

\subsection{Non-adiabatic quantum instanton}

It is clear from the above discussion that the PQI method is very closely related to Wolynes theory.\cite{Wolynes87} This connection can be made more explicit by noting that Wolynes theory and the PQI method can be regarded as the golden rule limit and adiabatic limit, respectively, of a more general non-adiabatic QI method. 

In order to derive this NAQI method one simply proceeds as in Eqs.~(27) to~(32), but with the flux operator defined as in Eq.~(6) with the generalised projection operator $\hat{P}_p(\alpha)$ in Eq.~(14).
The final result has the same form as Eq.~(31),
\begin{equation}
k_{\rm NAQI}Q_r = {1\over \beta^2\hbar}\sqrt{2\pi\over -\beta F_{\alpha^*}''(\lambda_{\rm sp})}e^{-\beta F_{\alpha^*}(\lambda_{\rm sp})},
\end{equation}
where
\begin{align}
e^{-\beta F_{\alpha}(\lambda)} &= (\beta\hbar)^2{\rm tr}_n\left[U_r(i(\beta-\lambda)\hbar)\,\hat{F}U_{p}(i\lambda\hbar)\,\hat{F}\right]\nonumber\\ &= (\beta\hbar)^2 c_{\alpha}(i(\lambda-\beta/2)\hbar)
\end{align}
and we have defined $\alpha^*$ and $\lambda_{\rm sp}$ as the values of $\alpha$ and $\lambda$ that maximise $F_{\alpha}(\lambda)$ and minimise $c_{\alpha}(i(\lambda-\beta/2)\hbar)$.

When $\Delta\to 0$ and $\alpha^*\to \pi/2$, as is the case in the golden rule limit, Eq.~(33) reduces to Wolynes theory,\cite{Wolynes87} and when the upper adiabatic electronic state becomes thermally inaccessible and $\alpha^*\to 0$, it reduces to the adiabatic PQI of Vaillant \emph{et al.}\cite{Vaillant19} The NAQI method is thus a generalisation of these pre-existing methods which can be applied to reactions that are intermediate between the two limiting regimes.

\section{Results and Discussion}\label{Results}

To illustrate the accuracy of the NAQI approach we shall consider a simple generalisation of the exponential curve crossing model from Sec.~\ref{Exact_Theories},
\begin{subequations}
\begin{equation}
V_0(q) = A e^{+q/L} 
\end{equation} 
\begin{equation}
V_1(q)=A e^{-q/L}-\epsilon,
\end{equation} 
\end{subequations}
which allows for a bias of $\epsilon$ towards the products.
Defining the crossing point of the two diabats as the solution to the equation $V_0(q^{\ddagger})=V_1(q^{\ddagger})$, and the value of the potential at the crossing point as $V^\ddagger=V_0(q^{\ddagger})$, the behaviour of this model can be fully characterised by the four dimensionless parameters, $\beta\Delta$, $\beta V^{\ddagger}$, $\beta\epsilon$ and $mL^2/\beta\hbar^2$. Note that $A=\sqrt{(V^{\ddagger}+\epsilon/2)^2-\epsilon^2/4}$.
 
To demonstrate the behaviour of the NAQI method in different regimes we shall consider three systems in which the values of $\beta V^{\ddagger}$, $\beta\epsilon$ and $mL^2/\beta\hbar^2$ are fixed while $\beta\Delta$ is varied so as to span the range from golden rule to Born-Oppenheimer like behaviour. For each value of $\beta\Delta$, the NAQI rate was evaluated using a sine finite basis representation (FBR) in the barrier region, with $\alpha$ and $\lambda$ optimised along with the location of the position space dividing surface, $s(q)=(q-q_0)/L$, so as to minimise $c_{\alpha}(i(\lambda-\beta/2)\hbar)$. The exact rate was computed for comparison by integrating the cumulative reaction probability
\begin{equation}
kQ_r\beta\hbar = \frac{1}{ 2\pi}\int_0^{\infty} e^{-\beta E}N(E)\,\beta{\rm d}E,
\end{equation}
with $N(E)$ calculated using the coupled channel log derivative method.\cite{Johnson73} 

In order to illustrate the importance of nuclear quantum effects, we shall compare the exact and NAQI results with the classical Born-Oppenheimer rate
\begin{equation}
k_{\text{cl-BO}} Q_r \beta\hbar= \frac{1}{2\pi}e^{-\beta U_-^{\ddagger}},
\end{equation}
where $U_{-}^{\ddagger}=\frac{1}{2}\Big(\sqrt{4(A-\Delta)^2+\epsilon^2}-\epsilon\Big)$ is the maximum on the lower adiabatic potential, and with the classical golden rule rate
\begin{equation}
k_{\text{cl-GR}}Q_r \beta\hbar = \frac{\Delta^2\sqrt{2\pi m\beta}}{\hbar|V_0'(q^\ddagger)-V_1'(q^\ddagger)|}e^{-\beta V^{\ddagger}},
\end{equation}
where $|V_0'(q^\ddagger)-V_1'(q^\ddagger)|=(2V^{\ddagger}+\epsilon)/L$. In terms of the four dimensionless parameters this is simply
\begin{equation}
k_{\text{cl-GR}}Q_r \beta\hbar = \frac{\beta^2\Delta^2}{(2\beta V^{\ddagger}+\beta\epsilon)} \sqrt{\frac{2\pi mL^2}{\beta\hbar^2}}e^{-\beta V^{\ddagger}}.
\end{equation}

\begin{figure}[t]
 \resizebox{0.8\columnwidth}{!} {\includegraphics{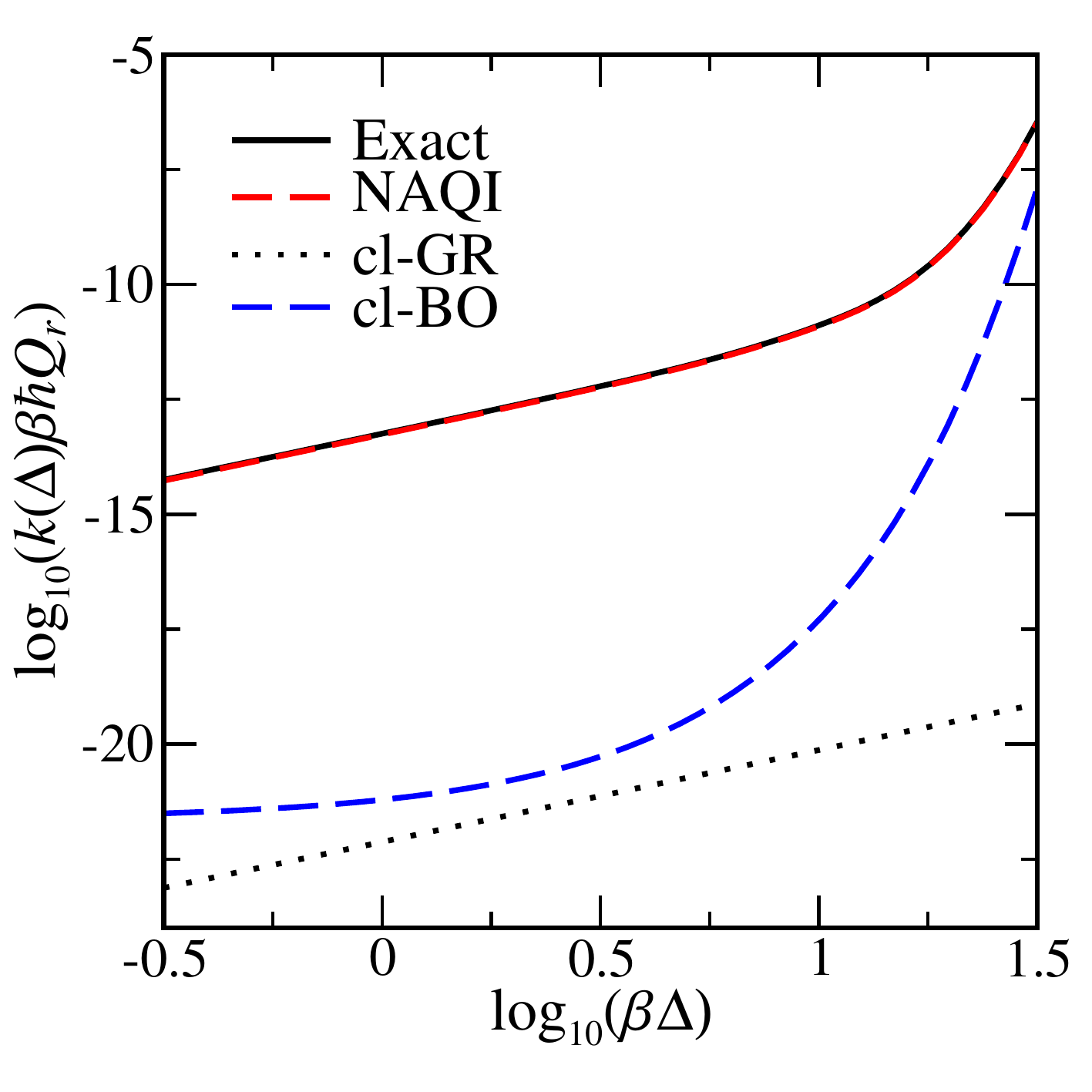}}
 \centering
 \caption{Comparison of the exact and NAQI rates as a function of the diabatic coupling $\log(\beta\Delta)$ for a symmetric exponential crossing model with $\beta\epsilon=0$, $\beta V^{\ddagger}=48$ and $mL^2/\beta\hbar^2=1/4$. The classical Born-Oppenheimer and golden rule rates are included to illustrate the importance of nuclear quantum effects.}
 \label{NAQI_Example1}
 \end{figure}

The first system we shall consider is the symmetric problem from Sec.~\ref{Exact_Theories}, with $\beta\epsilon=0$, $\beta V^{\ddagger}=48$ and $mL^2/\beta\hbar^2=1/4$. These parameters were chosen so that when $\log_{10}(\beta\Delta)=1.5$ the Born-Oppenheimer problem on the lower adiabatic potential is similar to the $300\,\mathrm{K}$ symmetric Eckart barrier problem considered by Vaillant \emph{et al}.\cite{Vaillant19} Figure~\ref{NAQI_Example1} compares the NAQI rate with the exact rate for this system. Excellent agreement is obtained for the full range of $\beta\Delta$ considered. The error in the NAQI rate is approximately independent of $\beta\Delta$, and is slightly less than 5\%. Comparison with the classical golden rule and Born-Oppenheimer rates highlights the importance of nuclear quantum effects, showing in particular that the effect of tunnelling on the rate is strongly dependent on the electronic coupling strength. The tunnelling enhancement of the rate ranges from a factor of 30 at the largest value of electronic coupling, $\log_{10}(\beta\Delta)=1.5$, to 12 orders of magnitude in the golden rule limit. Clearly there is very efficient tunnelling through the narrow, nearly cusped potential energy barrier at small values of the coupling.

It is interesting to note that the rate constant for the intermediate value of coupling considered in Fig.~3, $\log_{10}(\beta\Delta)=-0.25$, is actually very well described by the golden rule limit in this system. Hence we see that the electronic coupling at which the transition from Born-Oppenheimer to golden rule behaviour occurs is larger for the rate than the flux-flux correlation function. This results in the optimum value of $\alpha$ remaining close to the adiabatic value while the rate exhibits an approximately quadratic dependence on the electronic coupling characteristic of the golden rule limit. We do not expect this to be a general feature -- it occurs here because this is a symmetric problem in the deep tunnelling regime. In fact, we believe this helps to explain the success of ``mean field'' methods for such problems.\cite{Schwieters99,Duke16} These methods exploit the fact that at intermediate values of the electronic coupling the dominant contribution to the partition function in the barrier region is the $\Delta^2$ term. The $\Delta^0$ terms have a larger action due to lack of tunnelling whereas the $\Delta^2$ term is dominated by paths near the golden rule instanton which only spend a brief amount of time near the crossing point and have a smaller action. Note that the mean field instanton based methods all involve some kind of {\em ad hoc} approximation in which the $\Delta^0$ term is thrown away, whereas this is unnecessary in the present NAQI approach.

 \begin{figure}[t]
 \resizebox{0.8\columnwidth}{!} {\includegraphics{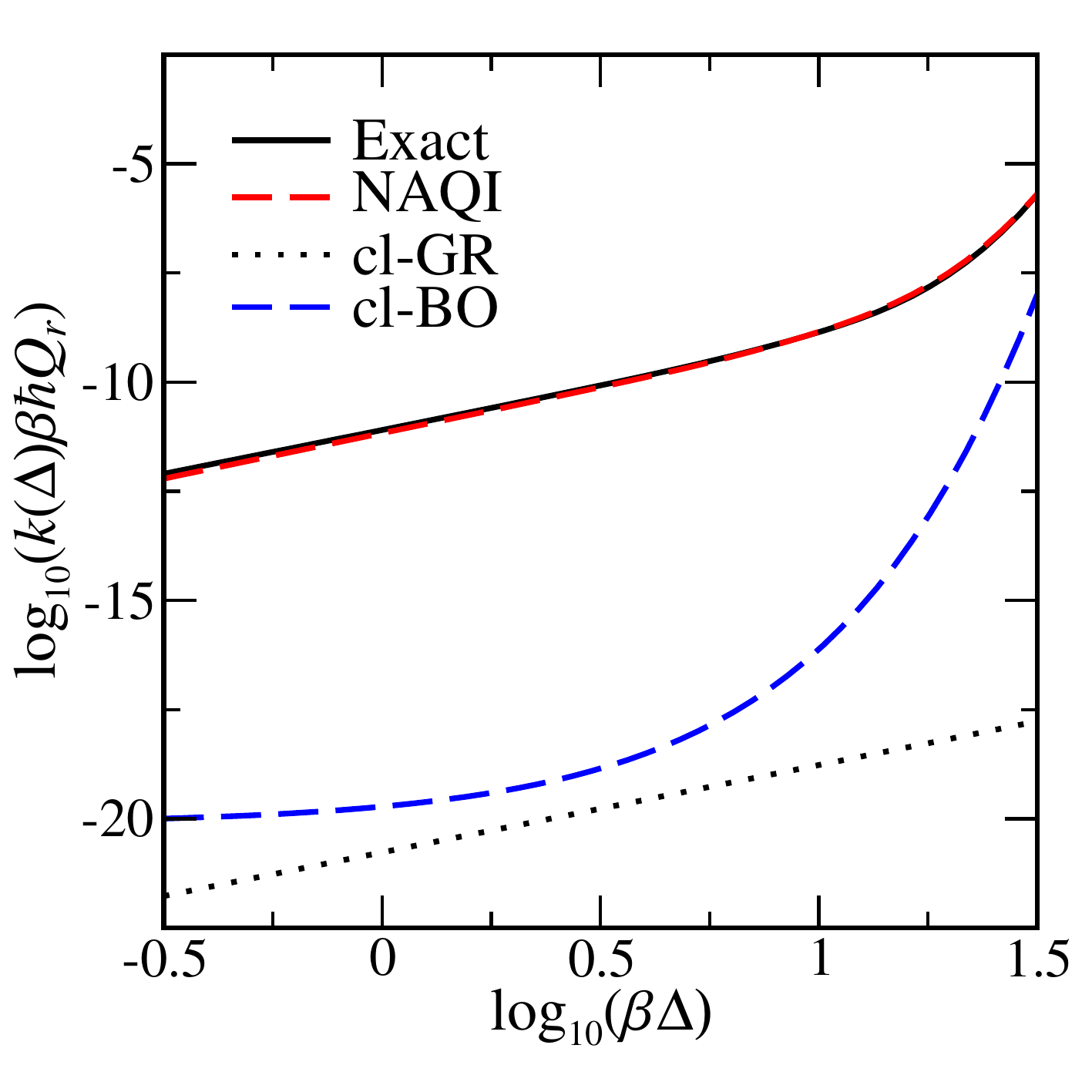}}
 \centering
 \caption{Comparison of the exact and NAQI rates as a function of the diabatic coupling $\log(\beta\Delta)$ for an asymmetric exponential crossing model with $\beta\epsilon=50$, $\beta V^{\ddagger}=44.5$ and $mL^2/\beta\hbar^2=16/49$. The classical Born-Oppenheimer and golden rule rates are included to illustrate the importance of nuclear quantum effects.}
 \label{NAQI_Example2}
 \end{figure}
 
  \begin{figure}[t]
 \resizebox{0.8\columnwidth}{!} {\includegraphics{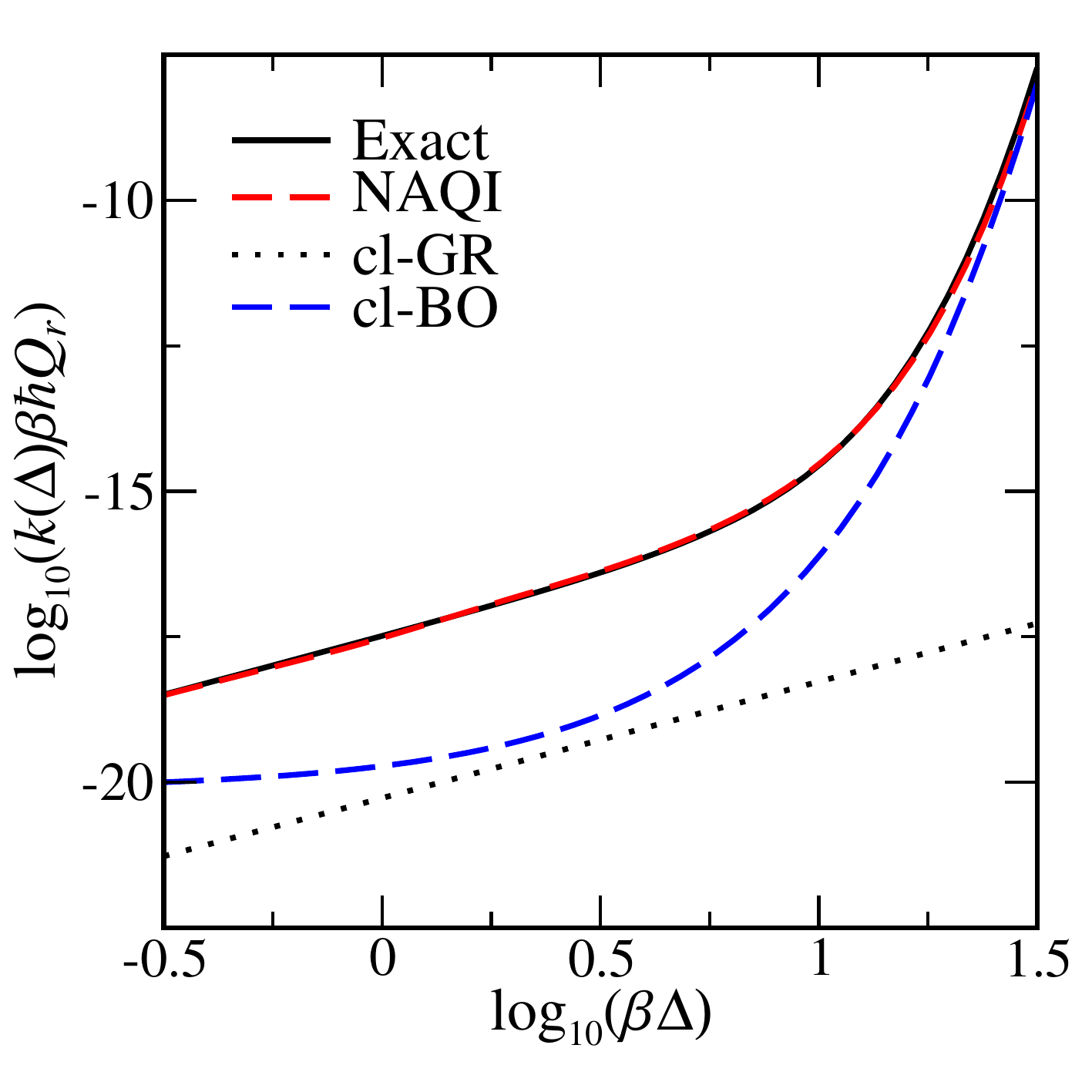}}
 \centering
 \caption{Comparison of the exact and NAQI rates as a function of the diabatic coupling $\log(\beta\Delta)$ for an asymmetric exponential crossing model with $\beta\epsilon=50$, $\beta V^{\ddagger}=44.5$ and $mL^2/\beta\hbar^2=160/49$. The classical Born-Oppenheimer and golden rule rates are included to illustrate the importance of nuclear quantum effects.}
 \label{NAQI_Example3}
 \end{figure}

Figure \ref{NAQI_Example2} shows the various rates as a function of the electronic coupling for the second system we shall consider, a strongly asymmetric system with $\beta\epsilon=50$, $\beta V^{\ddagger}=44.5$ and $mL^2/\beta\hbar^2=16/49$. Again the parameters were chosen such that when $\log(\beta\Delta)=1.5$ the lower adiabatic potential is similar to that in one of the systems studied by Vaillant \emph{et al.}, in this case the most asymmetric system with $\alpha=4$ at 300~K in Fig.~5 of Ref.~\onlinecite{Vaillant19}. As with the first system we see large nuclear quantum effects at all values of electronic coupling, with the largest quantum enhancement in the golden rule limit. We find excellent agreement between the exact rate and the NAQI rate, with the largest error near $\log(\beta\Delta)=-0.5$ where the NAQI rate underestimates the exact rate by around $20\%$. Interestingly the rate is again well described by Fermi's golden rule at this value of $\beta\Delta$ despite the correlation function and the optimum value of $\alpha$ being intermediate between the Born-Oppenheimer and golden rule regimes. However, in contrast to the symmetric system considered above, the optimum value of $\alpha$ moves significantly away from zero before the rate has begun to exhibit a quadratic dependence on the electronic coupling. 

The high accuracy of the method for such an asymmetric system at all values of the electronic coupling is particularly encouraging as it indicates that we are correctly capturing the instanton in this system. In particular, it shows that the introduction of the projection operators into the flux-flux correlation function works not only in the golden rule and adiabatic limits but also at intermediate values of the electronic coupling. In this regard the present method provides a significant improvement over previous theories, such as the QTST of Schwieters and Voth,\cite{Schwieters98,Schwieters99} which break down for strongly asymmetric systems.

The final system we shall consider is significantly less quantum mechanical while still being strongly asymmetric. This is achieved by increasing the dimensionless mass parameter by a factor of 10 relative to the previous model, so that $\beta\epsilon=50$, $\beta V^{\ddagger}=44.5$ and $mL^2/\beta\hbar^2=160/49$. Figure \ref{NAQI_Example3} compares the exact rate with the NAQI rate as well as the classical golden rule and Born-Oppenheimer rates for this system. We see that for the largest values of electronic coupling the classical adiabatic rate becomes a very good approximation to the exact rate, indicating that nuclear quantum effects are minimal in this regime. The NAQI rate again agrees very closely with exact rate for all values of the electronic coupling, with errors less than $5\%$ for $\log_{10}(\beta\Delta)<1$. The largest errors are observed at the largest coupling strengths, in the range $1.25\leq\log_{10}(\beta\Delta)\leq1.5$. We find that the NAQI approximation underestimates the exact rate by about $40\%$ at the upper end of this range, where the reaction is approximately classical and adiabatic. This is a well known deficiency of the adiabatic quantum instanton,\cite{Miller03,Vanicek05,Vaillant19} and arises because in the classical limit the correlation function is not well approximated by a Gaussian due to a long time polynomially decaying tail. Analysis of the free particle correlation function predicts that the adiabatic PQI underestimates the exact rate by $37\%$,\cite{Vaillant19} which is entirely consistent with the error seen here in the NAQI rate. Simple fixes have been suggested in the past to correct the adiabatic quantum instanton for this error,\cite{Miller03,Vanicek05} and it may be possible to apply similar fixes to the NAQI. While Wolynes theory, and hence the golden rule limit of NAQI, reduces to Marcus theory in the high temperature limit for the spin-boson model, it is known that they do not give the correct classical result at high temperatures for anharmonic systems. Recently however new methods have been suggested which aim to ameliorate this flaw whilst still accurately describing the low temperature regime,\cite{Thapa19,Fang19} and it may well be possible to generalise these approaches to arbitrary coupling in much the same way we have done here.

\section{Concluding Remarks}\label{conclusion}

We have demonstrated that it is straightforward to generalise the PQI approximation and Wolynes theory, which are applicable in the Born-Oppenheimer and golden rule limits respectively, to treat non-adiabatic reactions with arbitrary electronic coupling strengths. The resulting NAQI approximation has been shown to be highly accurate for both symmetric and strongly asymmetric systems at low temperatures where nuclear quantum effects are important. However, as is expected from its connection with the adiabatic quantum instanton, we find that it underestimates the exact rate by approximately $40\%$ in the adiabatic limit at high temperature, where the flux-flux correlation function is not well approximated by a Gaussian but instead exhibits a polynomially decaying tail. This problem has previously been overcome by assuming a different functional form for the correlation function or by going to a higher order asymptotic approximation, and these are both interesting avenues to explore to improve the present method.\cite{Miller03,Cao90,Cao95b,Yang05}

Here we have only applied the NAQI approach to simple one dimensional systems in order to demonstrate the basic features of the method. In order to apply the method to multidimensional systems it will be necessary to develop a path integral implementation, which we expect to be straightforward to do starting from the standard path integral implementation of Wolynes theory.\cite{Wolynes87,Lawrence18} One of the main difficulties associated with the calculation of multidimensional quantum instanton rates in the adiabtatic limit is locating the optimum position space dividing surface, and this difficulty will clearly carry over to the present method. We note however that the natural reaction coordinate in a two level system is expected to be the diabatic energy gap, and hence for many systems, especially those modelled with empirical valence bond force fields, finding an appropriate dividing surface may not be too much of an issue. (However the generalisation to systems with more than two electronic states will clearly be much more challenging). 

We have recently suggested an alternative approach to calculating non-adiabatic reaction rates which avoids the need to optimise the dividing surface.\cite{Lawrence19a,Lawrence20} The idea is to combine the Born-Oppenheimer and golden rule rates with an appropriate interpolation formula. When the Born-Oppenheimer rate is calculated using RPMD rate theory,\cite{Craig05a,Craig05b} and the golden rule rate using Wolynes theory,\cite{Wolynes87} the resulting interpolated rate is independent of the choice of position space dividing surface for all values of the electronic coupling strength.\cite{Lawrence19a} However the present NAQI approach clearly provides some advantages over the use of an interpolation formula. For example, one could imagine using its path integral implementation to obtain direct information about the imaginary time trajectories that are important in the reaction, and how these trajectories change as a function of the electronic coupling. 

Finally, we note that by making steepest descent approximations to the integrals over position in PQI and Wolynes theory one can obtain the semiclassical instanton in the Born-Oppenheimer and golden rule limits respectively.\cite{Richardson15b,Vaillant19} Hence, we expect that a steepest descent approximation to the NAQI should lead to an accurate semiclassical instanton which is valid for arbitrary electronic coupling strengths. More speculatively, because the semiclassical instanton in the adiabatic limit has a close connection to (and has been shown to provide an {\em a posteriori} justification for) RPMD rate theory,\cite{Richardson09} one might hope that the resulting non-adiabatic semiclassical instanton may help in the development of an accurate generalisation of RPMD for calculating non-adiabatic rates.

\section*{Data availability statement}

The data that support the finding of this study are available in the paper itself.

\begin{acknowledgments}
We would like to thank Jeremy Richardson and Manish Thapa for helpful discussions. J. E. Lawrence is supported by The Queen's College Cyril and Phillis Long Scholarship in conjunction with the Clarendon Fund of the University of Oxford and by the EPRSC Centre for Doctoral Training in Theory and Modelling in the Chemical Sciences, EPSRC grant no. EP/L015722/1. 
 \end{acknowledgments}
 
 \appendix
 
 \section{Diabatic projection operator}
 
 The alternative expression for the diabatic projection operator in Eq.~(12) can be derived by noting that
 \begin{equation}
 \theta(-\hat{\sigma}_z) = \lim_{\epsilon\to 0_+} {1\over 2\pi i}\int_{-\infty}^{\infty} {e^{-ix\hat{\sigma}_z}\over {x-i\epsilon}}\,{\rm d}x
 \end{equation}
 and
 \begin{equation}
 e^{-ix\hat{\sigma}_z} = e^{-ix}\dyad{0}{0}+e^{+ix}\dyad{1}{1}
 \end{equation}
immediately give 
 \begin{equation}
 \theta(-\hat{\sigma}_z) = \dyad{1}{1}.
 \end{equation}

\end{document}